# Long-living plasmoids generation by high-voltage discharge through thin conducting layers

A.L.Pirozerski[1] and S.E.Emelin[2]

[1] Research Institute of Physics, Saint-Petersburg State University, Russia
E-mail: *piroz@yandex.ru*

[2] Research Institute of Radiophysics, Saint-Petersburg State University, Russia
E-mail: *Sergei.Emelin@pobox.spbu.ru*

Abstract
A new type of pulse high voltage electric discharge through a thin conducting layer on the surface of glass plate has been investigated. The afterglow plasma of this discharge forms quasi-spherical object with a lifetime about 0.2–0.3 s. Electric properties of the objects were studied by electric probe method. Measurements of plasma radiation spectra kinetics at visible and near ultraviolet spectral ranges have been carried out. Comparative analysis of the physical properties of the plasmoids appearing in this discharges and of ones generated via thin metal wires burning is given. Possible mechanism of the plasma metastability are discussed.

### 1. Introduction

Recently a particular attention is attracted to autonomous long-living plasma formations arising at different types of erosional electric discharges which have a seeming resemblance with the ball lightning (BL) at beginning of their living, namely, a quasi-spherical form, average size about 10 cm and sufficiently bright luminescence. There exist several ways to generate similar objects, for example UHF discharge [1], electric explosion of a thin metal diaphragm **[2],** high-voltage erosive water discharge (HVEWD) [3,4], non-stationary discharge from a 380V power mains (RPMD) [5,6], burning of thin metal wires by electric discharge (MW) [5,7].

In [6] we have pointed out that these dust-gas fireballs have only partial (but possibly deep) similarity with the natural ball lightning and can be interesting only on three main aspects: the nature of energy storing, of radiation and shape stability of some possible BL types. In paper [4] we have revealed the physical nature of such formations as objects appearing due to structure-energy self-organization processes of chemically-active non-ideal plasma with smallest metal and/or dielectric particles.

Fireballs generation via a high-voltage discharge through thin conducting layers (TL) was proposed by us in [5]. The present paper is devoted to the further studies of physical properties TL plasmoids, as well as of the MW ones.

### 2. Experimental setup

The discharge circuit consists of a pulse storage capacitor 0.9 to 2.55 mF x 5kV, an inductance L= 40μH – 7.6mH integrated with pulse ignition transformer (resonance frequency 0.03 – 0.3MHz, pulse voltage up to 50kV), current limiting resistor (R=10–400 Ω), protecting spark gap (4 mm in length) and main discharger (TL or MW).

TL discharger had a dielectric base on which two poles were installed with a dielectric partition between them. To the each of the poles at height 100 mm narrow dielectric support was fastened with possibility to turn around parallel horizontal axes. Glass plates 100x25x2.4 mm with thin conducting $SnO_2$ layer and current-carrying electrodes (copper foil) placed above the layer were fixed on the supports by slats. Viewed from the side the glass plates had the form of the greek letter Λ. Resistance of the plates was varied in the range from 150 up to 300 Ω.

MW discharger had a similar construction, see [5].

For studying the discharge and the afterglow plasma spectrum dynamics we used an original spectrograph described in [6]. The spectrograph enabled to register the spectral distribution along the vertical direction at the spectral range 380–650 nm on a camcorder Sony DCR-TRV11E with the frame rate 50 field per second, the maximal spectral resolution being ~0.5 Å. As a reference source a He-Hg lamp with a reflector was installed behind the discharger. We carried out simultaneous video recording of the objects via another camcorder Sony DCR-HC30E.

Electric properties of the fireballs were studied via electric probe method. Several types of single and double probes were used. The probes were connected to a two-channel oscilloscope C1-83 through a resistive divider with input impedance 150 MΩ.

### 3. Experimental results.
### 3.1. General overview of the TL discharge.

The discharge processes and the fireballs evolution is similar to MW, RPMD or HVEWD ones. After discharge ignition a plasma cord appears, on the top of which a mushroom-like object formes. After current-breaking the object turns into a plasmoid of oval or of an irregular shape, see Figs. 1–3 below.

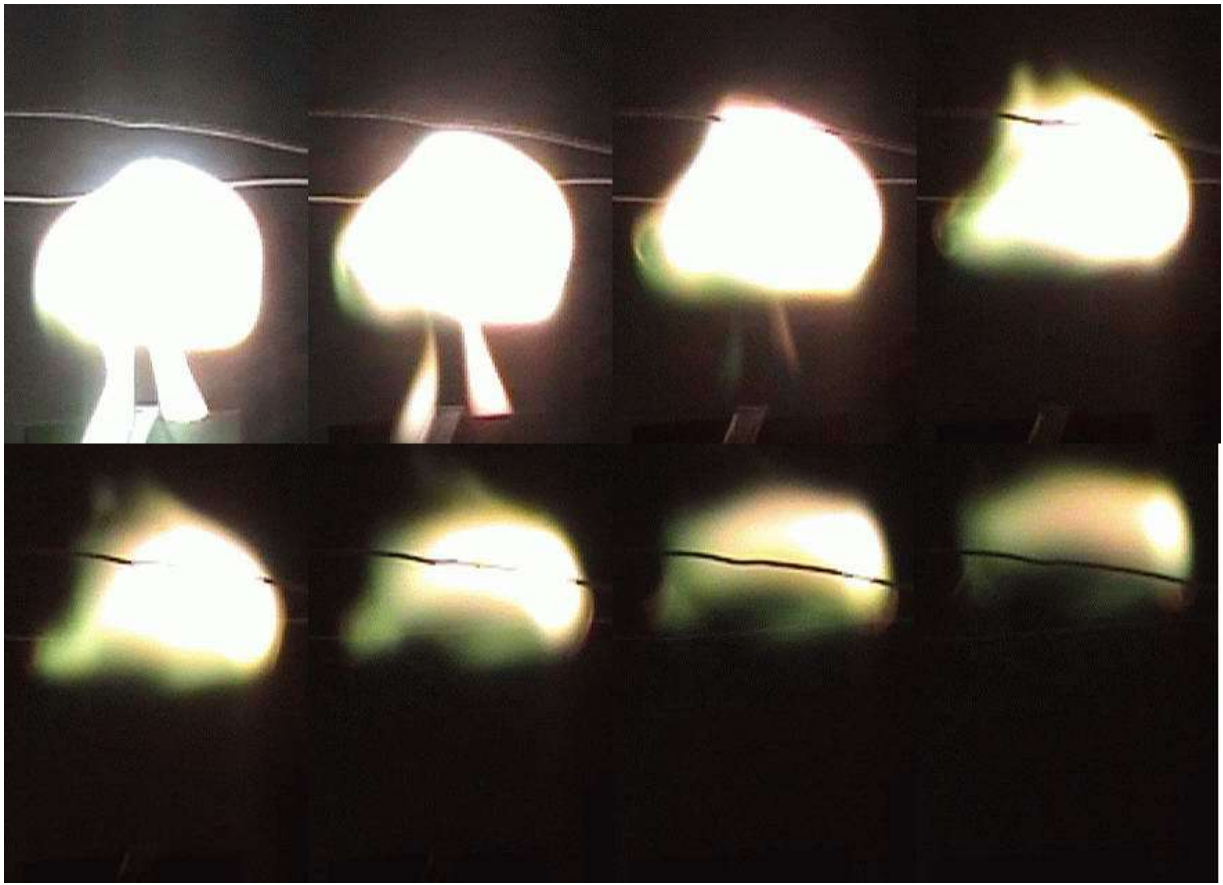

Fig.1. Evolution of a TL fireball. Time interval between frames is 20 ms. The current breaking occurs on the second frame (180 ms after the ignition). The wires are before and behind the object, it does not touch them.

It should be noted that the shape of plasmoids is strongly depends on some aerodynamic factors, primarily on presence and structure of airflows, disposition of surrounding objects, as well as on the ignition parameters and characteristic time of the discharge. At optimal combination of these factors enables to generate the plasmoids of a quasi-spherical shape, but for TL discharge it is sufficiently more difficult then for MW or RPMD ones. Also, no apparent correlation was found between shape and lifetime of TL fireballs. Relaxation of the objects can occur in two different ways. In most cases they turns into plasma torus (Figs. 1,2), their luminescent lasts ~ 0.2 s, while the residual dust toroid exists up to 1–2 s. Sometimes during the relaxation stage objects decreases in size with formation of a more bright oval area in the center, their shape becoming irregular.

Comparison of Fig.2 below and Fig.7 of paper [6] shows that surface tension of the TL objects is sufficiently weaker then in the case of RPMD ones, due to a lower concentration of dust particles.

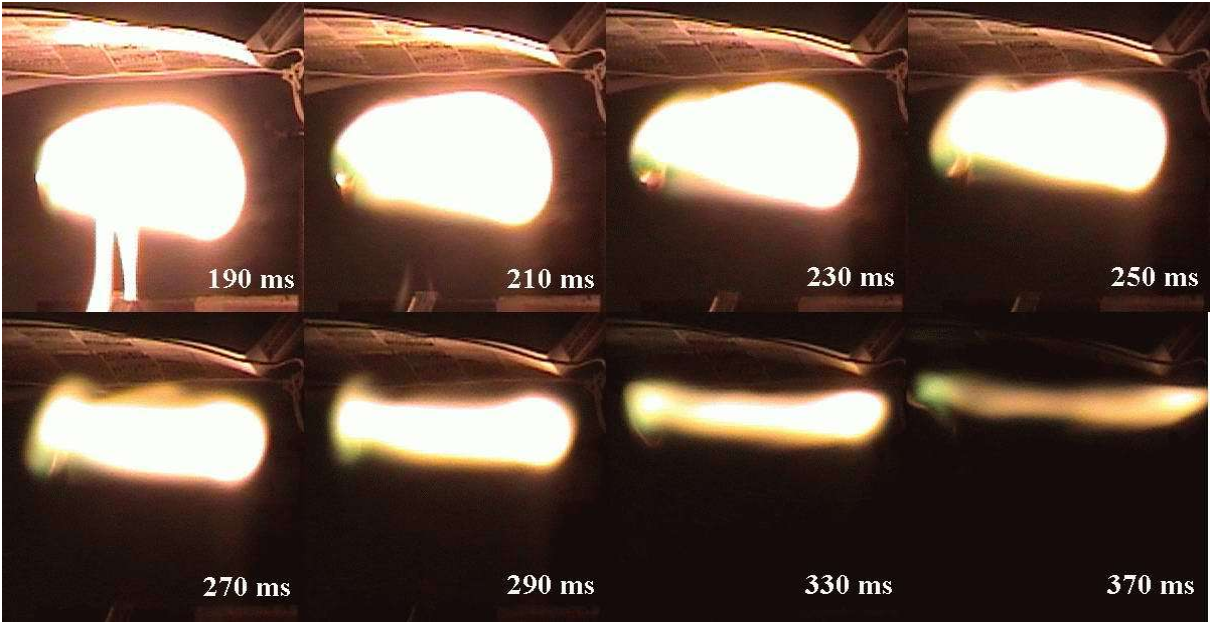

Fig.2. Relaxation of surface tension of a TL plasmoid. Indicated time is counted off from the discharge ignition.

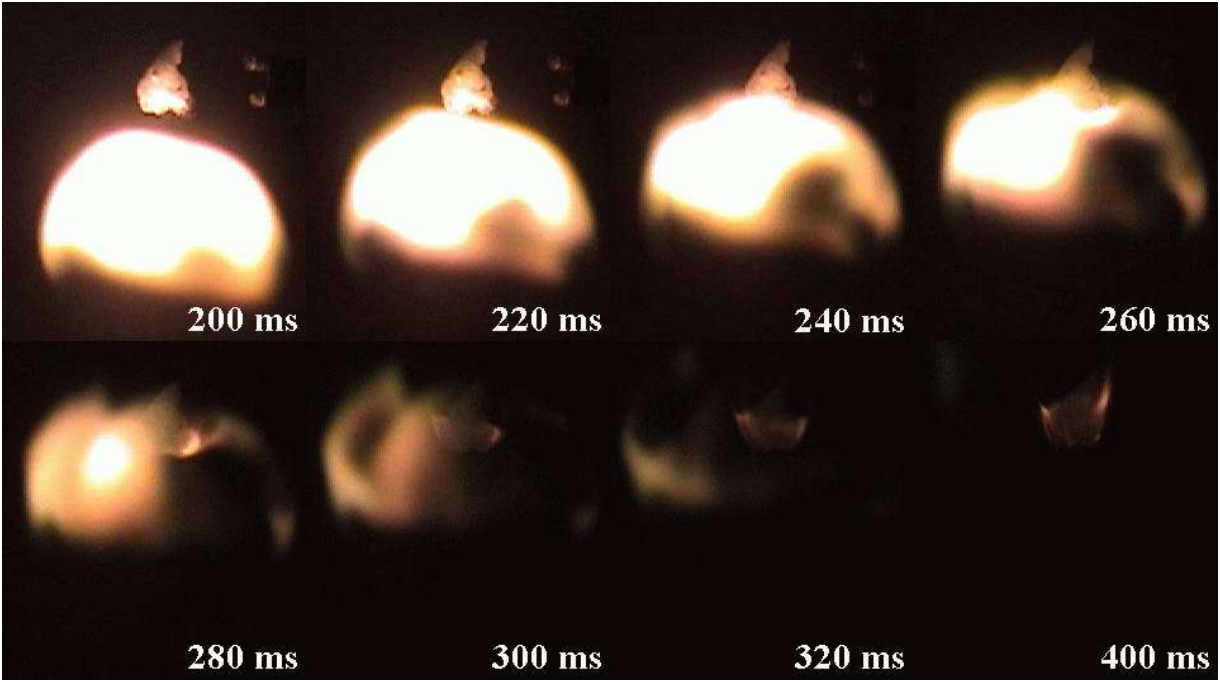

Fig. 3. Cotton wool firing by a TL plasmoid.

Similarly to RPMD, MW and HVEWD fireballs, at the beginning of the relaxation stage TL plasmoids are capable to set fire on cotton wool, but they can not set fire on a sheet of newsprint or burn through a thin aluminium foil.

### 3.2. Electric properties of TL and MW plasmoids

By evidences of eyewitnesses, the natural BLs have some unusual electric manifestations, which seems to be related with a large uncompensated electric charge. Our investigations shows that none of type of the objects (TL, MW, RPMD, HVEWD) have similar manifestation.

Firstly, it should be noted that at the same conditions the form and amplitude of probe signal were sufficiently varying from one experiment to another. It is most probably connected with low reproducibility of the shape, the internal gas-dynamics and the structure of fireballs and of its position relative to the probe.

For single probe measurements we used the following type of the probe: (a) a thin wire with uncovered end 3 mm in length, orientated vertically or horizontally; (b) a horizontal located copper foil plate 15x15 mm; (c) a horizontal located copper foil plate 100x100 mm; (d) the same plate but located vertically. Fig. 4 shows a signal from a TL fireball obtained with the probe (a). A pulse at ~150 ms corresponds to an interference at the moment of the current breaking, small sinusoidal components are noise from 50 Hz power main. The probe was located at the height 160 mm above the glasses tops. The sharp peak corresponds to the moment when the upper part of the object attained the probe (~25 ms after the current breaking, which is also confirmed by simultaneous video recording; so, the object can be considered as autonomous). The measured peak voltage was ~56 V. Taking into account that input impedance of the divider $R_d = 150$ M$\Omega$, we find by integrating the signal that total charge passing to the ground is about 3 nC. In general, the average charge was about 1.5 nC for TL objects and 2 nC for MW ones. Its polarity always corresponds to the one of the power supply ungrounded terminal.

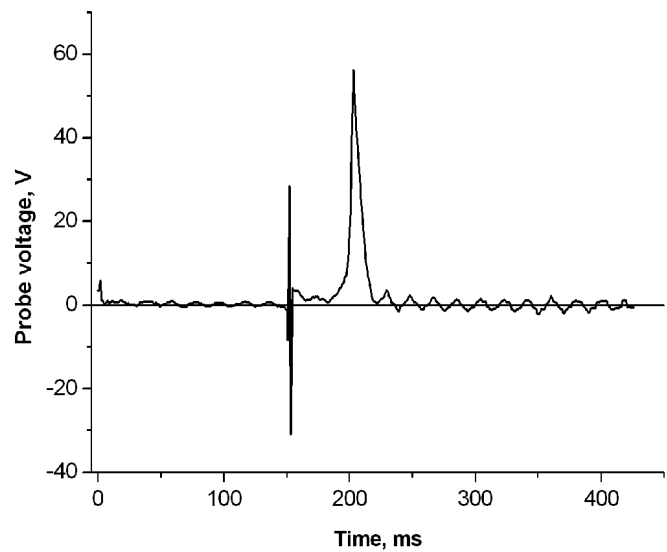

Fig. 4. Typical signal from a TL object registered with single electric probe.

In some cases two or three successive peaks were observed, which shows that only a part of the total charge of the plasmoid is taken by the probe. So, it is probable that the trailing edge of the pulse (which is an exponential decay with a characteristic time ~ 9 ms) is related to the conductivity decrease rather then to discharging of the object by the probe. The video recording gives the object vertical velocity ~0.75 m/s, hence the thickness of the conductive envelope is ~ 7 mm.

Assuming that the probe current is limited by the plasma conductivity we find total resistance of the discharge circuit is about $R_t$= 9 ms/6 pF = 1.5 GΩ >> $R_d$ , so the real plasma potential is 56 V * $R_t/R_d$=560 V, which is about a half of the storage capacitor voltage at the moment of the current breaking.

For MW objects, the use of probes (a), (b) and (d) give similar signals, their amplitude being generally 25% greater then for the probe (a) on TL objects. For TL objects no signals were registered with (d) probe, but amplitude of the 50 Hz noise increased in several times after the object had touched the probe and remained at this level during more than 250 ms (a similar but weaker effect can be seen even for (a) probe, see Fig. 4). Signal from probe (c) has low amplitude ( 8 – 12 V in the best cases) and longer duration (~ 45 ms and more). Total collected charge was ~2.5 nC in the best cases.

For double probe measurements we used a vertically located twisted pair cable with horizontally bent uncovered ends placed at different heights at the right angle. The height difference between the ends was varied from 3 to 20 mm. Each wire of the cable was connected to through identical resistive dividers to the corresponding channel of the oscilloscope operating at differential or alternate (switching frequency 500 KHz) mode.

For MW objects at the differential mode the signal was similar to one shown on Fig. 4. Measurements at the alternate mode shows that the upper probe gives very low signal, so the differential one was almost entirely due to the low probe.

So, we can conclude that both MW and TL objects obtain their charge from the discharge plasma cord; self-generated field of the objects, if present, is sufficiently lower.

### 3.3. Spectral characteristics of TL and MW discharges and objects

Spectra of TL and MW discharges and objects from UV to red spectral ranges are shown on the figures 5 – 20. When a discharge/objects spectrum was located on the image at sufficiently far from the reference one, a part of the image between them was cut out, on the corresponding figure the spectra are separated by a white line and reference lines are marked (Ref *) (otherwise – simply (Ref)). In this case especially a circular distortion should be taken into account when analyzing the spectra. Note that the time from discharge ignition when a spectrum was taken could be evaluated (by analyzing corresponding parallel videorecording of the discharge) only approximately, so the error can be ±10 ms and more, the same refer to the current breaking moment.

For TL the spectral range are divided into five overlapping subranges: 3700–4390 A, 4340–5040 A, 4980–5490 A, 5400–5920 A, 5850–6475 A. MW spectra were taken with a lower resolution and are divided into three subranges: 3735–4900 A, 4850–5920A, 5850–6410 A. Note that the camcorder sensitivity decreases rapidly from 3800 A to shorter wavelengths.

In paper [6] the presence of short-living and long-living components was found in the RPMD discharge radiation. For TL and MW ones the situation is similar. Fig. 5 shows the presence of a molecular bands CN 4216.0, 4197.2, 4181.0 in the TL discharge spectrum in ~30 ms after the ignition; a band at λ≤3890 is most likely also due to CN. They disappear in ~40 ms, see Fig. 6. Lines 3968 A and 3933 A belong most likely to Ca II ion (ionization energy 6.11 eV), i.e. 3968.47 ($E_{ex}$=3.12 eV) and 3933.67 (3.12 eV), but the presence of a residual Fe impurity cannot be totally excluded. Ca I lines, marked (1)-(6) on the Fig. 6, are 4318.65 (4.77 eV), 4307.74 (4.76 eV), 4302.53 (4.78 eV), 4298.99 (4.77 eV), 4289.36 (4.77 eV), 4283.01 (4.78 eV).

The line Ca I 4226.73 (2.93 eV) is present during the whole discharge; it is a unique atomic line at this range visible after the current breaking, see Fig. 7.

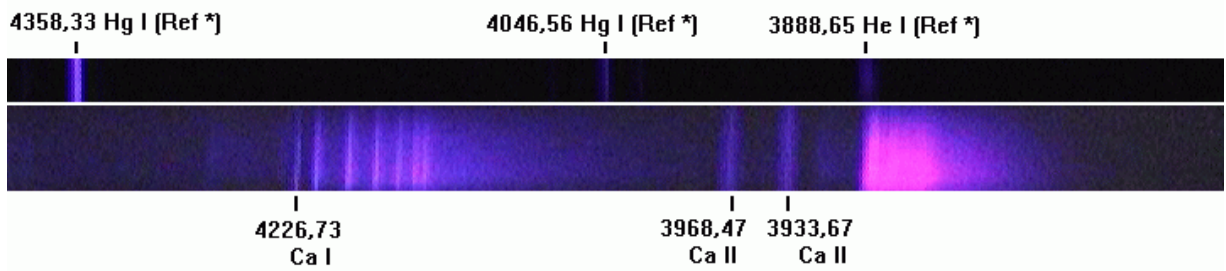

Fig. 5. TL discharge spectrum at UV–Blue range (3700 – 4390 A) in ~30 ms after the ignition.

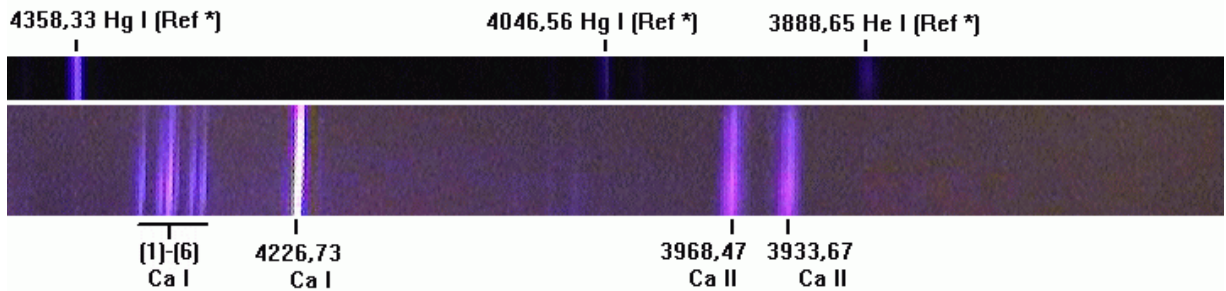

Fig. 6. TL discharge spectrum at UV–Blue range (3700 – 4390 A) in ~70 ms after the ignition.

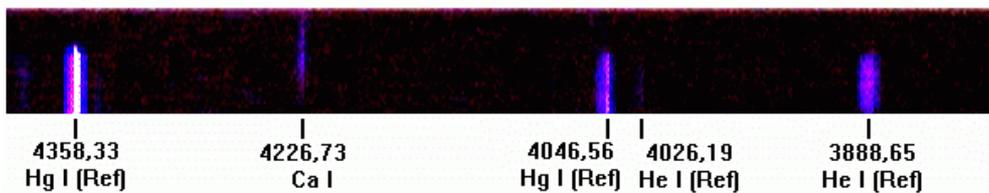

Fig. 7. TL object spectrum at UV–Blue range (3800 – 4390 A) just after the current breaking (~190 ms after the ignition). This image was processed to increase the brightness.

Radiation of both the discharge and object at blue-cyan range (4340–5040 A) is mainly due to AlO bands, but in bright lines of Ca I and Sr I are present, some of them can be seen still just after current breaking (see Figs. 8, 9).

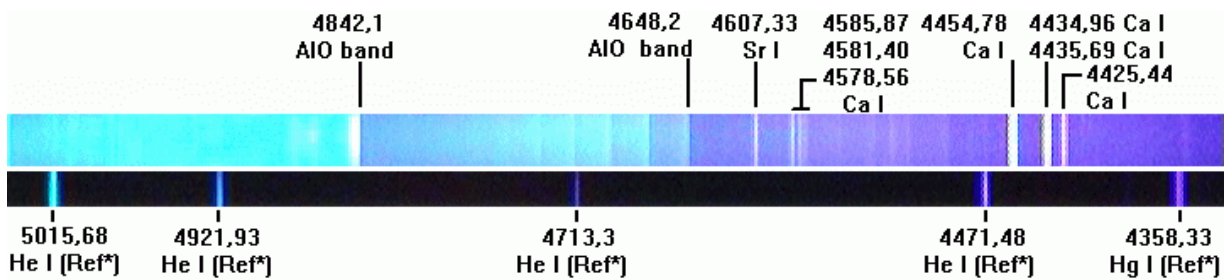

Fig. 8. TL discharge spectrum at Blue-Cyan range (4340 – 5040 A) in ~70 ms after the ignition.

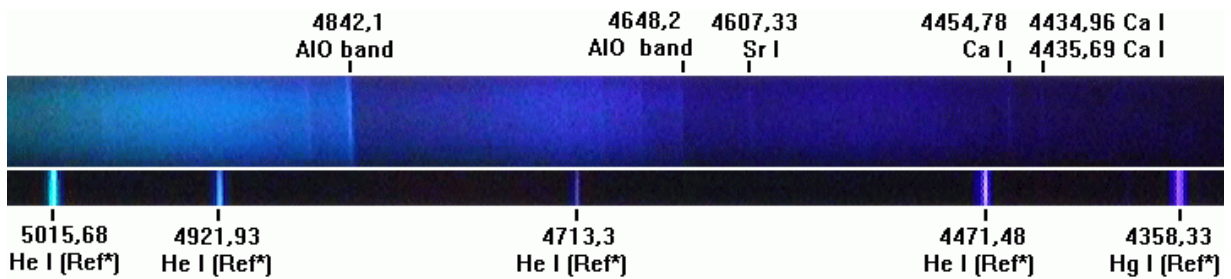

Fig. 9. TL object spectrum at Blue-Cyan range (4340 – 5040 A) just after the current breaking (~190 ms after the ignition).

The MW discharge spectrum at UV-Cyan range is rather different (Fig.10). Because of the lower resolution some lines and band can not be surely identified. Very bright lines Cu I 4062.7 (6.87 eV), 4063.29 (6.87 eV), Ca I 4226.73 and bands CaO 4221.9, NO$_2$ 4448.0 are present.

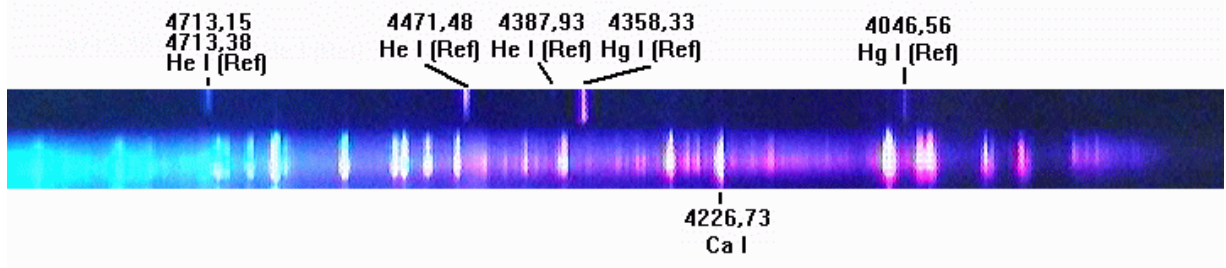

Fig. 10. MW discharge spectrum at UV–Cyan range (3735 – 4900 A) in ~60 ms after the ignition.

The MW object spectrum (Fig. 11) includes several unidentified bands at 4470–4790 (possibly, due to Cu$_2$O) and the line Ca I 4226.73.

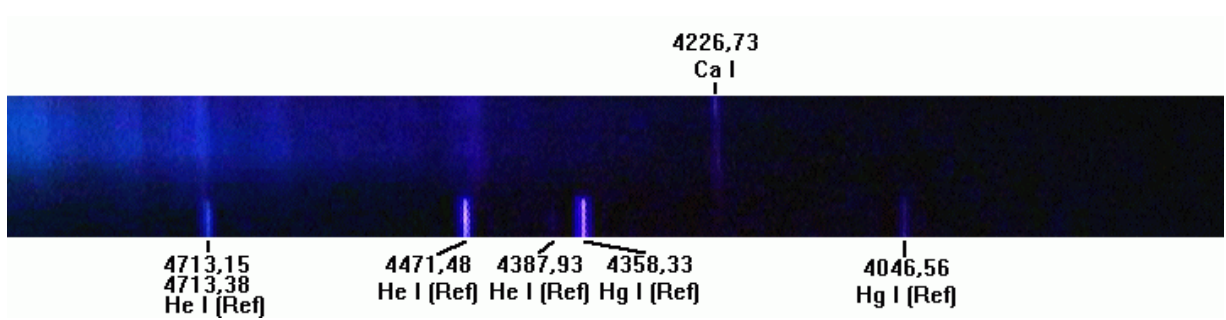

. Fig.11. MW object spectrum at UV–Cyan range (3735 – 4900 A) ~ 10 ms after the current breaking (~200 ms after the ignition).

At the cyan-yellow range (4890–5920) the TL discharge spectrum (Figs. 12, 14) is formed by Cu I lines 5105.54 (3.82 eV), 5153.24 and 5218.20 ( 6.19 eV), 5292.52 (7.74 eV), 5200.87 (7.8 eV), 5700.24 (3.82 eV), 5782.13 (3.79 eV), Mg I lines 5183.61, 5172.68, 5167.33 (5.11 eV), and Na I lines 5889.95 (2.11 eV), 5895.92 (2.10 eV) together with weaker molecular bands which are possibly due to CuO and Cu$_2$O. In the afterglow spectrum (Figs. 13, 15) just after current breaking the line Cu I 5105.54 (3.82 eV) is visible, but then only Na I lines remains.

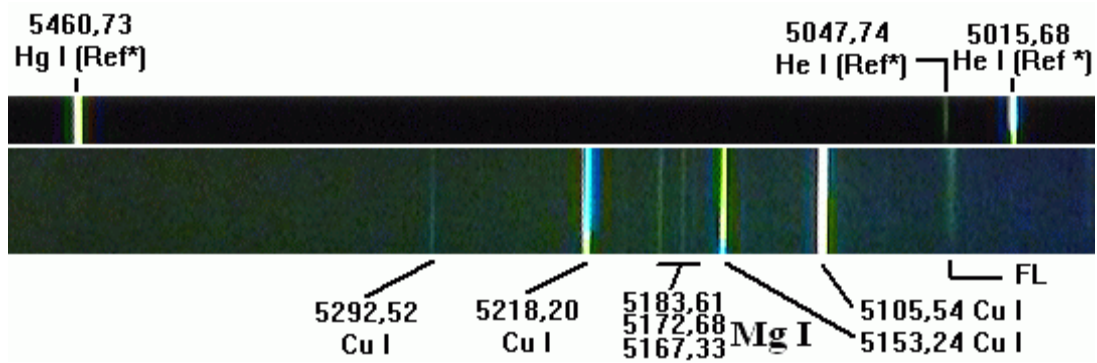

Fig. 12. TL discharge spectrum at Cyan-Green range (4980 – 5490 A) in ~90 ms after the ignition. A "line" marked FL is a optical flare from Cu I 5218.20 line.

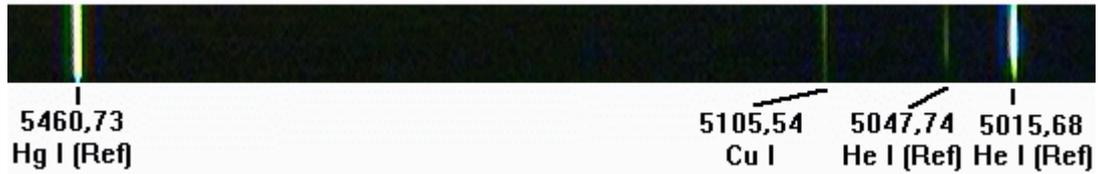

Fig. 13. TL object spectrum at Cyan-Green range (4980 – 5490 A) just after the current breaking (~170 ms after the ignition).

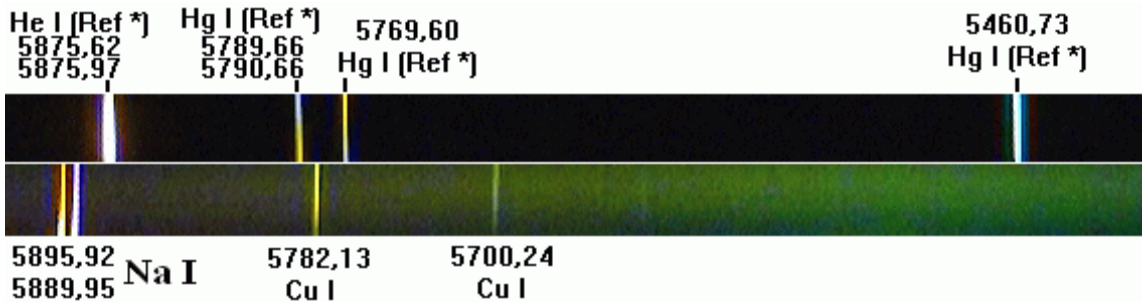

Fig. 14. TL discharge spectrum at Green-Yellow range (5400 – 5920 A) in ~40 ms after the ignition.

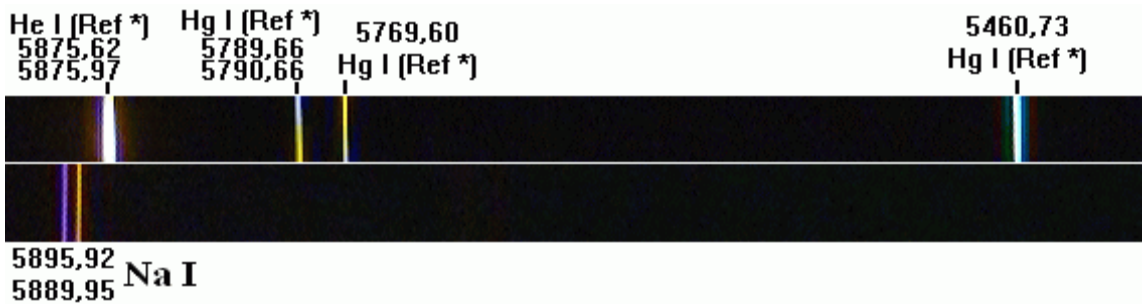

Fig. 15. TL object spectrum at Green-Yellow range (5400 – 5920 A) in ~10 ms after the current breaking (~180 ms after the ignition).

The MW discharge and afterglow have sufficiently higher intensity at this spectral range, see Fig. 16, 17. In the afterglow several Cu I lines are visible as well as molecular bands belonging most probably to CuO and $Cu_2O$.

At the red-yellow range TL and MW discharges and afterglows have similar spectra (see Figs. 18–20), but MW ones have higher intensity. The afterglow is mainly due to Na I lines and CuO bands.

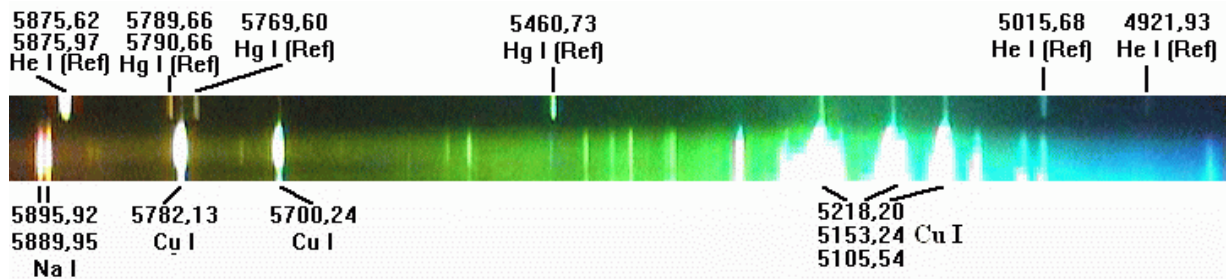

Fig. 16. MW discharge spectrum at Cyan-Yellow range (4850 – 5920 A) in ~50 ms after the ignition.

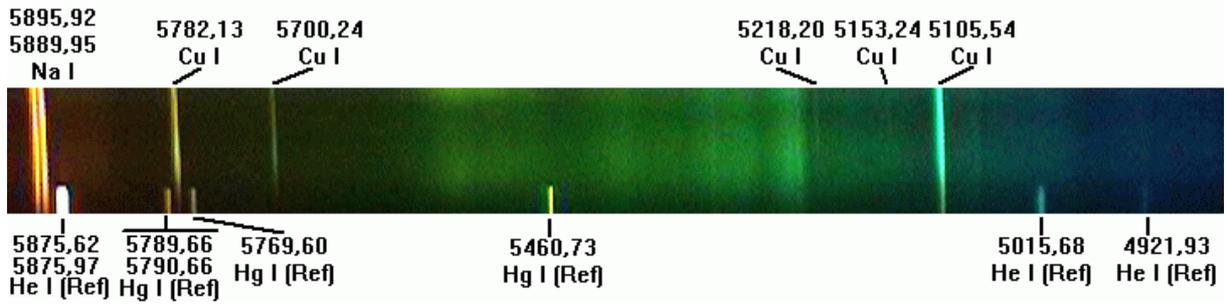

Fig. 17. MW object spectrum at Cyan-Yellow range (4850 – 5920 A) in ~20 ms after the current breaking (~170 ms after the ignition).

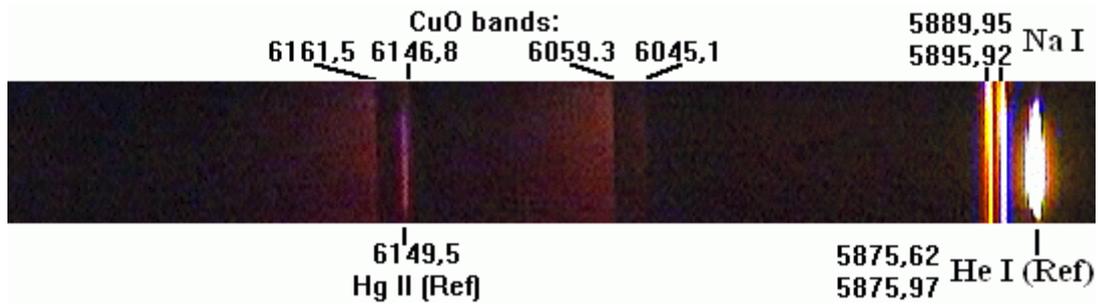

Fig. 18. TL object spectrum at Yellow-Red range (5850 – 6475 A) just after the current breaking (~170 ms after the ignition).

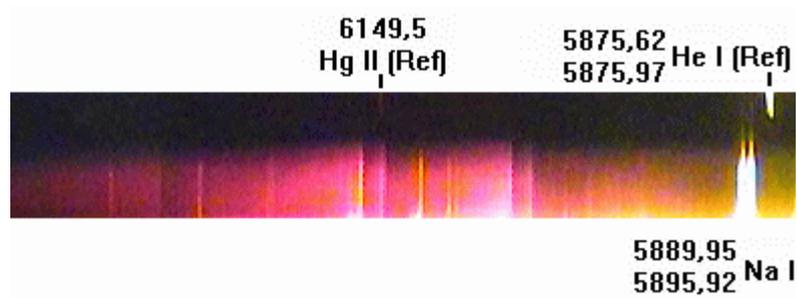

Fig. 19. MW discharge spectrum at Yellow-Red range (5850 – 6410 A) in ~30 ms after the ignition.

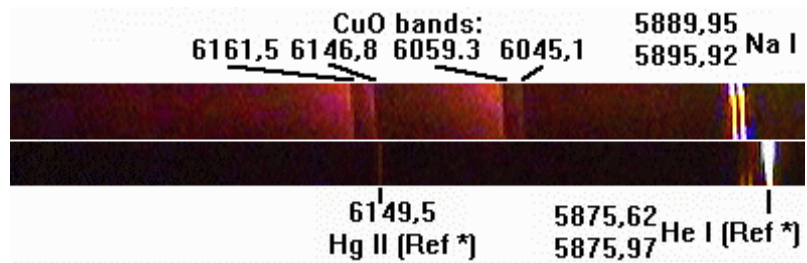

Fig. 20. MW object spectrum at yellow-red range (5850 – 6410 A) in ~10 ms after the current breaking (~210 ms after the ignition).

4. Conclusions

The results of our researches presented in this article confirm main ideas stated in our paper [6]. Summarizing them and also taking into account results of [5,7], we arrive at the following conclusions:

- all laboratory fireballs, i.e. MW, TL, RPMD and HVEWD objects have only partial similarity to the natural phenomenon;
- the afterglow luminescence has a non-thermal, non-equilibrium and collective character; the metastability of the objects is supported by a catalytic decay of chemically active plasma created during the discharge;
- the objects obtain their charge from the discharge plasma cord, their self-generated electric field, if present, is sufficiently weaker;
- presence or absence of an electric charge of the objects, as well as proximity their shape to spherical one, have no apparent influence on the afterglow duration.